# Ultrafast Heat Flow in Heterostructures of Au Nanoclusters on Thin-Films: Atomic-Disorder Induced by Hot Electrons

*Thomas Vasileiadis†\*, Lutz Waldecker†¹, Dawn Foster‡, Alessandra Da Silva‡, Daniela Zahn†, Roman Bertoni†², Richard E. Palmer§ and Ralph Ernstorfer†\**

† Fritz-Haber-Institut, Faradayweg 4-6, 14195 Berlin, Germany

‡Nanoscale Physics Research Laboratory, School of Physics and Astronomy, University of Birmingham, Edgbaston, Birmingham B15 2TT, United Kingdom

§College of Engineering, Swansea University, Bay Campus, Fabian Way, Swansea SA1 8EN, United Kingdom

¹current address: Department of Applied Physics, Stanford University, Stanford, USA
²current address: University of Rennes, Institut de Physique de Rennes, UMR UR1-CNRS 6251, F-35000 Rennes, France

*E-mail: vasileiadis@fhi-berlin.mpg.de, ernstorfer@fhi-berlin.mpg.de

**ABSTRACT:**

We study the ultrafast structural dynamics, in response to electronic excitations, in heterostructures composed of size-selected Au nanoclusters on thin-film substrates with the use of femtosecond electron diffraction. Various forms of atomic motion, such as thermal vibrations, thermal expansion and lattice disordering, manifest as distinct and quantifiable reciprocal-space observables. In photo-excited, supported nanoclusters thermal equilibration proceeds through intrinsic heat flow, between their electrons and their lattice, and extrinsic heat flow between the nanoclusters and their substrate. For an in-depth understanding of this process, we have extended the two-temperature model to the case of 0D/2D heterostructures and used it to describe energy flow among the various subsystems, to quantify interfacial coupling constants, and to elucidate the role of the optical and thermal substrate properties. When lattice heating of Au nanoclusters is dominated by intrinsic heat flow, a reversible disordering of atomic positions occurs, which is absent when heat is injected as hot substrate-phonons. The present analysis indicates that hot electrons can distort the lattice of nanoclusters, even if the lattice temperature is below the equilibrium threshold for surface pre-melting. Based on simple considerations, the effect is interpreted as activation of surface diffusion due to modifications of the potential energy surface at high electronic temperatures. We discuss the implications of such a process in structural changes during surface chemical reactions.

GRAPHICAL ABSTRACT:

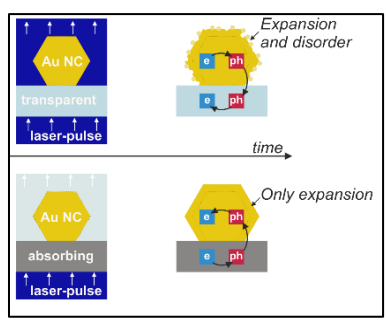

The most crucial achievement of nanotechnology, in line with Feynman's original vision, [1,2] is the precise arrangement of atoms to construct well-defined nanoscale building blocks. Today, the ability to synthesize metallic nanoclusters (NCs) of specific size and symmetry[3,4] and to deposit them on various surfaces with controlled density,[5] offers one possibility to optimize the numerous functionalities of confined metallic systems. Recent advances in the synthesis of size-selected NCs show that the procedure can be scaled up, in order to address realistic industrial needs.[6] From the perspective of basic science, being able to prepare size-selected metallic NCs is also very promising, since it allows detailed studies of their size-dependent optical,[7,8] electronic,[9] structural[10] and catalytic[11] properties.

In all cases, it is valuable to know under which circumstances the appealing symmetry of size-selected NCs becomes fragile; a serious limitation that was raised already at the early stages of nanotechnology. Zero-dimensional metallic systems are known to have a strong tendency towards amorphization compared with their bulk analogs. For instance, the melting point of Au NCs decreases as the diameter becomes smaller.[12] In conditions close to equilibrium, the stability of NCs has been examined using electron microscopy techniques, for various atomic structures[13] and temperatures,[14] while the nature of their equilibrium ground state has been explored using irradiation with electrons.[15] The situation is different in conditions far from equilibrium, where different subsystems of a solid can have drastically different temperatures for a very short period. Such non-equilibrium states are the key factor in numerous functionalities of metallic nanostructures, for instance photo-catalysis,[16] light-harvesting technologies[17] or photo-induced charge-transport.[18,19] To access the underlying interactions, energy flow and transient structural properties, it is necessary to employ time-resolved techniques that are sensitive to the structural symmetry.

Ultrafast changes of the lattice-order can be observed in a direct way using femtosecond diffraction.[20,21] More precisely, femtosecond electron diffraction (FED) is an appropriate investigatory tool for spatially confined systems like nanostructures and ultrathin films due to the high scattering cross-section of electrons. An in-depth understanding of the experimental results requires a realistic model of out-of-equilibrium thermodynamics. Particularly for heterostructures of low-dimensional materials, it is essential to consider not only the intrinsic properties of each component but also interactions across their interface.

In the present work, FED in combination with a model of heat flow in low-dimensional heterostructures, is used to study the lattice dynamics of photo-excited, size-selected Au$_{923\pm23}$ NCs on different substrates. Au NCs and other nanostructures have been studied with time-resolved diffraction in the past, in the melting and pre-melting regime using high incident laser fluences[22,23] as well as order-disorder dynamics of organic ligand/ nanoparticle supracrystals.[24] Clark *et al.*[25] have visualized surface premelting in a laser-excited Au nanocrystals composed of approximately $10^6$ atoms using coherent diffraction imaging. The participation of the substrate in heat flow has been discussed in the case of Au islands and nanoparticles on graphene,[26] while in a recent work Sokolowski-Tinten *et al.* have studied heat flow in heterostructures of Au thin-films and insulators.[27] Other zero-dimensional systems, examined with FED, were Bi nanoparticles[28] and quantum dots of GaAs[29] and PbSe.[30] Plech *et al.* and Hartland *et al.* observed signatures of surface melting of Au nanoparticles with optical spectroscopy.[31,32]

The experiments and modelling reported here aim to advance the field by acquiring a detailed picture of ultrafast heat flow in nanoscale heterostructures. On this basis, we demonstrate how different pathways can achieve similar maximum lattice temperatures but exhibit different time-evolution and different structural changes. We study the anharmonicity and lattice expansion of Au NCs and provide experimental evidence that lattice heating by hot electrons triggers ultrafast disordering of the Au NCs atomic lattice.

**RESULTS AND DISCUSSION:**

The samples under investigation consist of homogeneously dispersed, size-selected clusters composed of 923±23 gold atoms (from now on noted as $Au_{923}$ for simplicity). The $Au_{923}$ NCs are supported on thin films of 20 nm thick amorphous carbon (a-C) or 10 nm thick silicon nitride membranes (Si-N), see Methods for details. Aberration-corrected Scanning Transmission Electron Microscopy (STEM) in High-Angle Annular Dark-Field (HAADF) mode was used to characterize the distribution of NCs, a typical image of which is shown in **figure 1.a**. Most NCs were identified as isolated $Au_{923}$ (75%), while some aggregated and formed mostly dimers (25%). The deposited density was 8 NCs per 100 $nm^2$. The preparation conditions allowed, to a good extent, for adjusting the balance of the different structural allotropes.[3] In the present experiments, the amorphous-like Icosahedral structures (Ih) have been nearly eliminated and most of the NCs could be identified as Decahedral (Dh) or FCC (**fig. 1b**). In the FED experiments depicted in **figure 1c**, the samples were excited by a sub-100 femtosecond laser pulse with a central wavelength of 400 nm, resulting in electronic excitation of the NCs and/or the substrate by optical absorption. The response of the lattice is observed by recording diffraction patterns with ultrashort electron bunches at different time delays relative to the initiating optical excitation.[33]

The recorded diffraction patterns of $Au_{923}$ consisted of Debye-Scherrer rings, due to the random orientation of the NCs and the amorphous structure of both substrates. The diffraction patterns were radially averaged and the diffraction intensity was plotted as function of scattering angle. To extract the lattice dynamics, the diffraction peaks were fitted using pseudo-Voigt peak-profiles, after subtraction of a background that corresponds to inelastic scattering and electrons scattered from the substrate. The selected function for the background contained the diffraction pattern of a bare substrate recorded at the same conditions, a Lorentzian function for the tail of the zero-order peak and finally a third-order polynomial fitted in the flat regions between the various diffraction peaks (**fig. S1**). The result of background subtraction and fitting is shown in **figure 1.d** for $Au_{923}$ on a-C and the corresponding diffraction pattern is shown as an inset. The extracted intensities, positions and widths of diffraction peaks for different pump-probe delays reveal phononic excitations, lattice expansion and atomic disordering after photo-excitation (**fig. 1.e**).

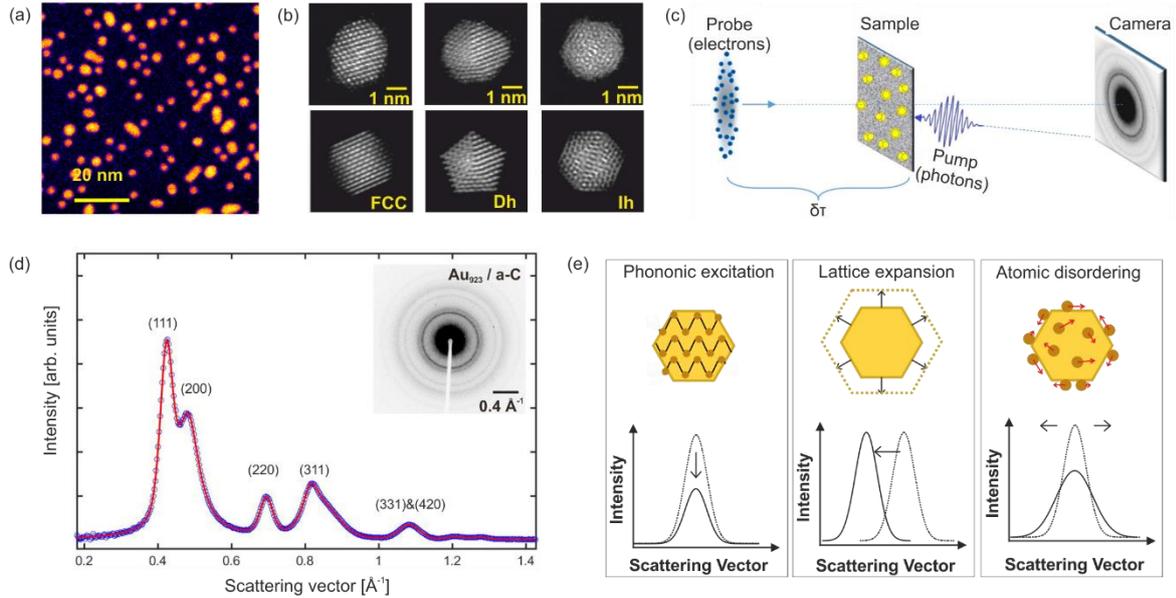

**Figure 1: Electron microscopy and diffraction of size-selected $Au_{923}$ NCs.**
**(a)** The distribution of $Au_{923}$ NCs on the surface of a-C as observed by STEM-HAADF imaging. **(b)** High resolution experimental images (upper image) and the corresponding electron-scattering simulations (lower image) of individual NCs. **(c)** Schematic illustration of the FED experiment. A femtosecond laser pulse arrives at the sample at nearly normal incidence and initiates dynamics (pump). The response of the lattice is probed using short electron bunches (probe) that arrive at selected time-delays (δτ). **(d)** Radial average of diffraction signal for $Au_{923}$ NCs on a-C (blue circles) and fitting using pseudo-Voigt profiles (red line) after background subtraction. The inset shows the raw recorded diffraction pattern. **(e)** The observables of FED and the corresponding atomic motions in real space.

**Heat flow dynamics in low-dimensional heterostructures:** After optical excitation, the electrons thermalize within tens of femtoseconds[34] and begin to transfer heat to the lattice of $Au_{923}$ through electron-phonon coupling. For full restoration of thermodynamic equilibrium there should be, in addition to electron-phonon coupling, exchange of heat between the NCs and the substrate materials, through electronic or vibrational coupling. All the above result in a time-varying energy content of the NCs lattice. The accompanying structural changes become evident as changes in the intensity, position and width of the diffraction peaks. As the lattice temperature of $Au_{923}$ NCs increases, the intensities of the diffraction peaks decay and the background of inelastically scattered electrons increases due to the Debye-Waller effect. Furthermore, the increased lattice temperature coupled with the anharmonicity of the interatomic potential leads to thermal expansion, which becomes evident as a shift of diffraction peaks to lower-scattering vector. This observable would be diminished in the case of thin films where lattice expansion is constrained in two dimensions. Finally any atomic motion that leads to a deterioration of crystallinity results in peak-broadening. Because of these effects, FED becomes sensitive to changes of lattice temperature and can be used to study the dynamics of nanoscale heat flow.

Previous FED experiments on bulk-like Au thin-films have measured a time-constant for full electron-lattice equilibration of ~4-6 ps[35,36] followed by thermal relaxation of the sample on the ns-μs time scale. The observed dynamics of supported $Au_{923}$ NCs are very different than bulk Au and differ markedly between the two substrates we employed. In this case, all observables have bi-exponential behavior at pump-probe delays shorter than 1 ns (for the underlying formulation see **Sup. Inf. 2**). The decay of diffraction peak intensity and the increase of the background (Debye-Waller effect) can be seen in **figure 2.a** for a-C and **figure 2.b** for Si-N supports. For a-C the time-constants are $\tau_1=(6\pm2)$ ps for the fast process and $\tau_2=(80\pm40)$ ps for the slow. For Si-N as substrate, the fast process has a time-constant of $\tau_1=(5.0\pm0.7)$ ps and the slow one, which exhibits now the reverse effect, a time-constant of $\tau_2=(120\pm50)$ ps. The average thermal expansion is calculated straightforwardly from the position of diffraction peaks and is shown for both substrates in **figure 2.c**. The dynamics of expansion are similar with that of the Debye-Waller effect.

For both substrates, the NC coverage was the same and so the different dynamics are due to the different optical and thermal properties of the substrate materials. In the case of a-C, both substrate and NCs are directly excited by the laser pulse. In contrast, Si-N remains largely unaffected since it is transparent at this wavelength. Thus, the equilibration with the substrate proceeds through heating of $Au_{923}$ on a-C and cooling of $Au_{923}$ on Si-N. Based on the lattice dynamics observed in bulk Au and the dependence of the slow process on the optical properties of the substrate, the fast process can be attributed to intrinsic electron-phonon interaction and the slow process to extrinsic coupling to the substrate, which can be used for quantitative studies of interfacial heat transport.[37] The relative amplitudes of the bi-exponential fitting suggest that extrinsic heating of $Au_{923}$ NCs by heat flow from a-C contributes ~60% to the overall effect. Full thermal relaxation back to the room temperature ground state occurs on the ns-μs timescale, *i.e.*, before the arrival of the next pump pulse, and is not investigated in this work.

To gain access into the thermodynamic properties of $Au_{923}$ and to enable a quantitative analysis of the microscopic energy flow, it is useful to extract the atomic mean-square-displacement (MSD) from the peak decay using the Debye-Waller relationship:[38]

$$\frac{I_{hkl}(t)}{I_{hkl}(t<0)} = \exp\left[-\frac{4\pi^2}{3D_{hkl}^2}(<u^2>_{(t)} - <u^2>_{(t<0)})\right] \quad (1)$$

This direct conversion between intensity ($I_{hkl}$) and MSD ($<u^2>$) is possible because the experiments are performed in the single scattering regime. One important detail is that it is preferable to use the total intensity and not simply the height of the diffraction peaks in the analysis, since in the latter case, possible changes of the peak profile shape are mistreated as peak decay due to the Debye-Waller effect. The values for the interplanar spacing ($D_{hkl}$) in the above equation evolve with time because of the cluster expansion, as described below. With knowledge of the temperature dependence of the MSD,[39] it is now possible to extract the lattice temperature evolution as effective measure of the time-dependent energy content of the lattice.

We point out that the lattice temperature in general is an ill-defined quantity in strong non-equilibrium states after intense photo-excitation, since different phonon modes can have different electron-phonon coupling strengths leading to non-thermal phonon distributions.[40,41] These effects can be taken into account either with the support of *ab initio* molecular dynamics[40,42] or through a momentum-resolved measurement of phonon dynamics.[41,43] Both tasks are currently unfeasible for large ensembles of randomly oriented NCs with hundreds of atoms. We therefore adopt the two-temperature approximation in the present work.

The estimated lattice temperature, plotted together with the observed relative expansion (**fig. 2.d**), can be used to extract the thermal expansion-coefficient. Within the experimental noise and the accuracy of the fitting procedure, the different peaks ((111), (220) and the average of (331) and (420)) give all the same expansion coefficient (**fig. 2.d**). One of the basic assumptions of using the temperature-dependence of MSD to estimate the lattice temperature is that lattice vibrations are harmonic. For $Au_{923}$, anharmonic contributions in the Debye-Waller factor do not cause significant deviations since all peaks give similar evolution of temperature and expansion.

In line with this finding, the measured thermal expansion coefficient appears to be suppressed in comparison with bulk Au. For the temperature range reached by ultrafast lattice heating the average thermal expansion coefficient of $Au_{923}$ is $a_L=(9.1\pm0.3)\cdot10^{-6}$ $K^{-1}$ on a-C and $a_L=(9.9\pm0.6)\cdot10^{-6}$ $K^{-1}$ on Si-N . For comparison, **figure 2.d** also contains a linear increase with a slope of $14\cdot10^{-6}$ $K^{-1}$, which is the thermal expansion coefficient of bulk Au at room temperature.[44] Spatial confinement is thus reducing thermal expansion by ~30%. This trend has been observed in the past, for similar sizes of Au NCs, using EXAFS,[45] but also for other elements such as Pt nanoparticles[46] and it has been attributed to the large contribution of surface tension. This surface-induced compression of the Au nano-lattice, can be causing the increased frequencies of certain vibrational modes of NCs compared to bulk Au.[47] To confirm that surface-tension compresses Au$_{923}$ NCs, we have deposited them on few-layer-graphene and recorded static diffraction patterns. By comparing the scattering vector of the (220) peak of Au ($D_{220}$ = 1.4391 Å, for bulk Au) with the (110) peak of graphene ($D_{110}$ = 1.2280 Å) we obtain a contraction of (0.63±0.05)% (**Sup. Info. 3**). The bond-length contraction, measured with EXAFS for similar sizes,[48,49] was in the order of ~0.5-2%.

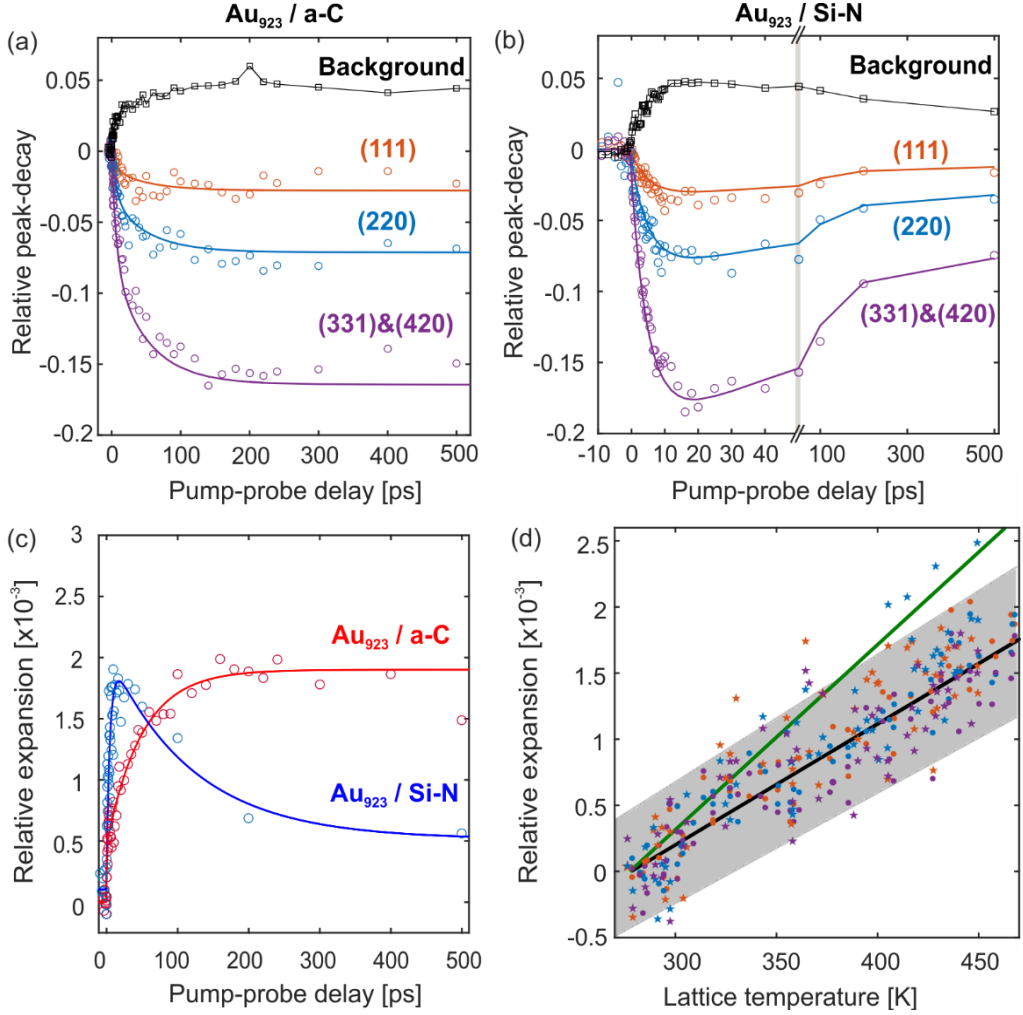

**Figure 2: Laser-induced Debye-Waller effect and expansion dynamics of Au$_{923}$ / thin-film heterostructures. (a)** Peak-decay and background-increase, as a function of pump-probe delay time, for three diffraction peaks of Au$_{923}$ on a-C (open circles) and predicted evolution based on the Debye-Waller relationship (solid lines). **(b)** Plot of the same quantities for Au$_{923}$ on Si-N. **(c)** Relative expansion of the NCs as a function of time (open circles) and bi-exponential fitting (solid lines) for Au$_{923}$ on a-C (red) and on Si-N (blue). **(d)** Plot of relative expansion as a function of temperature for the (111), (220) and the average of (331) and (420) peaks of Au$_{923}$ on a-C (orange, blue and magenta filled circles) and the average over the same peaks for Au$_{923}$ on Si-N (same color filled stars). Each data-point corresponds to a different pump-probe delay

and hence different, effective lattice temperatures. The solid black line is a linear fit that is used to extract the expansion coefficient. The grey area shows the 95% confidence intervals for a-C. Most data-points of Si-N are located within these limits. The green solid line shows, for comparison, the expected slope based on the thermal expansion coefficient of bulk Au at room temperature.

**Intrinsic and extrinsic heat flow in Au$_{923}$ NCs:** The dynamics of heat flow, in bulk-like, metallic thin films, have been studied extensively by means of the so-called two-temperature-model[50–52] (TTM). In this model, the electrons and the lattice are treated as two separate heat baths in thermal contact, each one having its own heat capacity and temperature. The incident laser pulse is expressed with a source term in the equation for the electronic temperature and the energy transfer rate from the hot electrons to the cold lattice through the electron-phonon coupling constant.

The observed substrate-dependent dynamics can be used to conceive the most basic extension of the two-temperature model in the case of low-dimensional heterostructures, a schematic illustration of which is shown in **figure 3.a**. This extended model contains one source term for each electronic component of the heterostructure, since the substrate can also absorb light. Heat flow is then intrinsic, between the electrons and the lattice of each material, and extrinsic, between the two different materials. Extrinsic heat flow can be carried out by transmission of hot electrons or phonons through the interface.

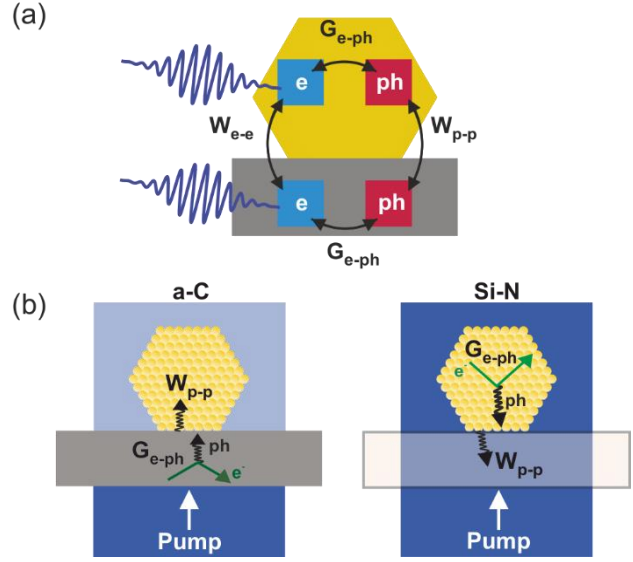

**Figure 3: Schematic illustration of the model of heat flow and the observables of FED.** (a) Four-temperature-model for NC / thin-film heterostructures. The arrows indicate the couplings between different subsystems of the heterostructure: $G_{e-ph}$ for electron-phonon coupling within each component, $W_{e-e}$ for electronic coupling with the substrate and $W_{p-p}$ for vibrational coupling with the substrate. (b) Different pathways for lattice-heating of the NCs: extrinsic heating from the laser-excited substrate (left) and intrinsic heating through electron-phonon coupling for a transparent substrate (right).

The simple schematic of **figure 3.a** can be translated into four coupled differential equations that describe the time evolving energy contents of the four involved subsystems. These coupled differential equations can be used for a non-linear fit of the experimental lattice temperature evolution and have the form:

$$\frac{dT_E^{Au}}{dt} = -\frac{G_{E-PH}^{Au}}{C_E^{Au}}(T_E^{Au} - T_L^{Au}) + \frac{W_{e-e}}{L^{Au}C_E^{Au}}(T_E^S - T_E^{Au}) + S^{Au}(t) \quad (2)$$

$$\frac{dT_L^{Au}}{dt} = \frac{G_{E-PH}^{Au}}{C_L^{Au}}(T_E^{Au} - T_L^{Au}) + \frac{W_{p-p}}{L^{Au}C_E^{Au}}(T_L^S - T_L^{Au}) \quad (3)$$

$$\frac{dT_E^S}{dt} = -\frac{G_{E-PH}^S}{C_E^C}(T_E^S - T_L^S) - \frac{W_{e-e}}{L^C C_E^C}(T_E^S - T_E^{Au}) + S^S(t) \quad (4)$$

$$\frac{dT_L^S}{dt} = \frac{G_{E-PH}^S}{C_L^C}(T_E^S - T_L^S) - \frac{W_{p-p}}{L^C C_E^{Au}}(T_L^S - T_L^{Au}). \quad (5)$$

The variables T and C represent the temperature and the heat capacity of each subsystem. The upper indices denote the material (Au for $Au_{923}$ or S for substrate) and the lower indices the subsystem (E for electrons and L for lattice). The electron-phonon coupling constants are denoted by $G_{e-ph}$ and the interfacial electronic and vibrational couplings are expressed by $W_{e-e}$ and $W_{p-p}$. Finally, the relative masses of the two materials are included in the equations through the effective thickness of the NCs-layer ($L_{Au}$=1.4 nm) and the thickness of the substrate ($L_S$). The last terms of equations (3) and (5) express the temporal evolution of the absorbed laser pulse and are given by:

$$S(t) = \frac{F}{L} \frac{\exp[-4\ln(2)(t-t_o)^2/w^2]}{w/2\sqrt{\pi/\ln(2)}}, \qquad (6)$$

where w is the full-width half maximum duration of the laser pulse (w = 50 fs), F the absorbed fluence, L the thickness and $C_E$ the electronic heat capacity of each material. The time resolution of the experimental setup is taken into account by convoluting the temperature evolutions by a Gaussian located at $t_o$.

The free parameters of the fitting are: (i) the absorbed fluences, $F_{Au}$ and $F_S$, (ii) the extrinsic coupling constants $W_{e-e}$, $W_{p-p}$ and (iii) the intrinsic electron-phonon coupling constant $G_{e-ph}$ for $Au_{923}$. For the electron heat capacity of $Au_{923}$ we have used the electron heat capacity of bulk Au since electron confinement effects are expected to play a minor role for this number of atoms due to thermal broadening and screening. The lattice heat capacity of $Au_{923}$ has been estimated by the works of Sauceda et al.[53,54] (**see also Sup. Info. 4**).

In the absence of any experimental data for the electron-phonon coupling constant of a-C, its value has been selected in order to fit the reported values for the time-constant of electron-lattice interactions in graphite.[55] The preparation method of the a-C thin-films ensures that C atoms have an $sp^2$ character and hence the heat capacity was approximated by literature values of graphite.[56] A more detailed description of intrinsic interactions in graphitic substrates could be done, for instance, by splitting the lattice heat bath in more reservoirs, to separate the so-called strongly coupled optical phonons (SCOPS) from all other modes.[57–59] To achieve a defined initial excitation, we have excited the sample from the side of a-C. Illumination from the side of the NCs results in an undefined distribution of the laser energy due to plasmonic enhancement of the light-matter interaction caused by the Au NCs.[60]

The fitted curves, together with the experimental evolutions of the lattice temperature, are plotted in **figure 4.a** for both substrates. The maximum increase of lattice temperature is similar for both substrates (~170 K) but the heat flow mechanism is very different. From fitting of the Au$_{923}$ on Si-N data we extract an electron-phonon coupling constant $G^{Au}_{e-ph}=(1.9\pm0.5)\cdot10^{16}$ W/m$^3$K and a vibrational coupling constant $W_{p-p} = (16\pm8)$ MW/m$^2$K. The data fitting reveals the extrinsic electronic coupling $W_{e-e}$ is insignificant compared to the other coupling constants, as expected for this metal-dielectric heterostructure. To improve fit convergence, we therefore repeated the analysis with $W_{e-e}$ set to zero. The electron-phonon coupling constant extracted from four measurements is $G^{Au}_{e-ph}=(2\pm0.2)\cdot10^{16}$ W/m$^3$K (**fig. S4**). The electron-phonon coupling constant has also been measured for free-standing thin films of Au with a thickness of 8 nm using the same apparatus, the Debye-Waller effect and the TTM (**fig. S5**). The measured value is $G_{bulk}=2.7\cdot10^{16}$ W/m$^3$K, in excellent agreement with optical pump-probe experiments analyzed with the TTM[61] that gave $G_{bulk}=2.61\cdot10^{16}$ W/m$^3$K.

The extracted electron-phonon coupling constant of Au$_{923}$ is ~70% of the value of bulk Au. Recent studies have shown that in the two-temperature approximation employed here the effect of transient non-thermal phonon distributions is understimated.[40-42,62] It is thus possible that the difference in the coupling constants reflects a weakening of phonon-phonon interactions in spatially confined systems and not a genuine reduction of the electron-phonon coupling strength. The observed reduction of the thermal expansion coefficient indicates that, on average, anharmonicity is reduced in metallic NCs. It is then expected that phonon-phonon interactions will be suppressed and the full thermalization of the lattice will be slower compared to bulk Au.

For Au$_{923}$ on a-C, both intrinsic and extrinsic heat flow contribute to lattice heating and the fitting becomes uncertain regarding $G^{Au}_{e\text{-}ph}$. We extract an electron-phonon coupling constant $G^{Au}_{e\text{-}ph}=(1.6\pm0.9)\cdot10^{16}$ W/m$^3$K. The vibrational coupling constant ($W_{p\text{-}p}$) was found to be $(90\pm10)$ MW/m$^2$K while, the electronic coupling was smaller by a factor of 10 and thus played a minor role. The vibrational coupling to a-C is larger by a factor of 5.6 compared with Si-N. Besides the selection rules of interfacial phonon-phonon interactions this quantity can also be affected by the surface morphology of the thin-film substrates. Thin-films of a-C are expected to contain a variety of graphitic nanostructures and hence to have higher surface roughness and binding area.

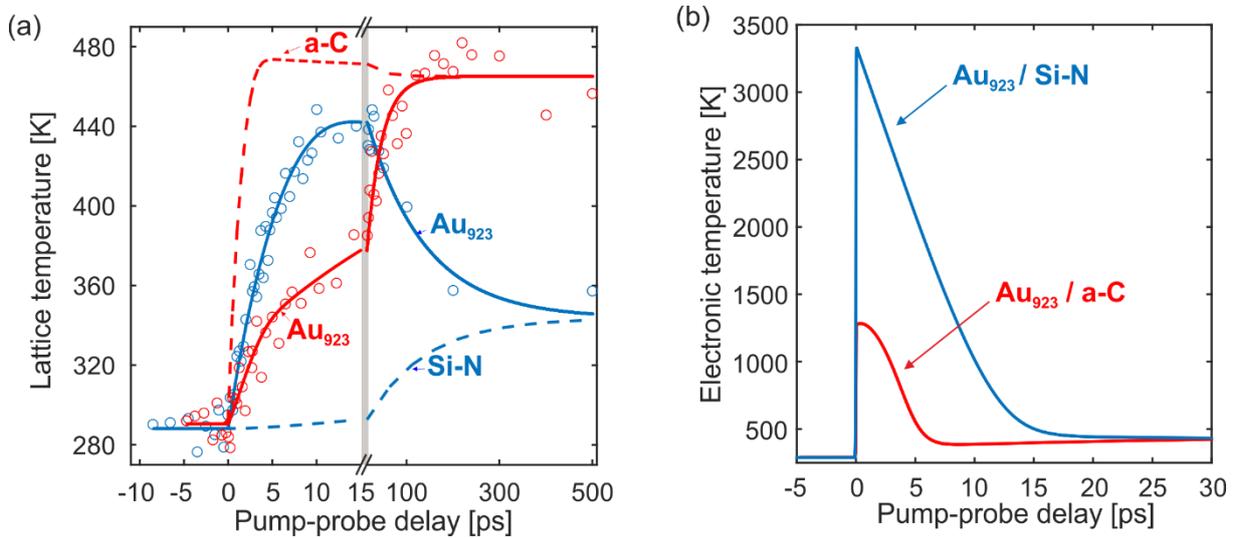

**Figure 4: Evolution of electronic and lattice temperature for absorbing and transparent substrates and comparison of disordering effects. (a)** Experimental evolution (open circles) of the Au NC lattice temperature on a-C (red) and on Si-N (blue, same color code in all subplots) and fitting based on the model of heat flow (solid lines). The predicted lattice temperature evolutions for the corresponding substrates are shown with dashed lines. **(b)** Prediction of the electronic temperature evolution at short time-delays for both substrates (same colors). The incident fluence was ~2.7 mJ/cm$^2$ and ~5.1 mJ/cm$^2$ for Si-N and a-C respectively.

**Structural changes accompanying energy flow:** Based on the model it is now possible to compare different pathways of heat flow and elucidate their role on the observed structural changes. The starting point is the predicted evolution of the electronic temperature in the NCs. The experiment on a-C has been carried out with an incident fluence of ~5.1 mJ/cm$^2$, and the experiment on Si-N with ~2.7 mJ/cm$^2$. In the first case, the electrons reach a relatively small maximum temperature of ~1300 K, because the absorbing substrate depletes the incoming pulse before it will arrive on the NCs. In contrast, for Au$_{923}$ on Si-N, the maximum electronic temperature is ~3300 K although the incident laser fluence has been reduced.

It has been shown (**fig. 2.c**) that the increased lattice energy content after photo-excitation is accompanied by lattice expansion and that the thermal expansion coefficient is similar for both types of substrate although energy flows in very distinct ways. However, the evolution of the peak width shows a dependence on the substrate properties, meaning the exact pathway of energy flow. In the case of extrinsic lattice heating of the NCs through vibrational coupling to the laser excited-substrate (Au$_{923}$ on a-C), the widths of the (111), (220), (331) and (420) peaks stay constant after photo-excitation (**fig. 5.a**). When lattice-heating is carried out solely by hot electrons (Au$_{923}$ on Si-N), the same diffraction peaks broaden after photo-excitation (**fig. 5.b**).

The average width increase is fitted with a bi-exponential function shown with a black, solid line in **figure 5.b**. The dynamics of the broadening are fast, comparable with the evolution of the lattice temperature (**fig. 4.a**). The rise in peak width can be described with an exponential rise with a time-constant of $\tau_1$=5.8±1.5 ps. This time constant is very similar with the time-constant of the Debye-Waller dynamics (**fig. 2.b**) or equivalently the lattice temperature increase (**fig. 4.a**). The maximum increase of the fast process is (2.8±0.4)% and subsequently the width decreases with the time constant $\tau_2$~200 ps back to the ground state value. Based on the bi-exponential fitting, the rise of the fast process begins upon the pulse arrival and not at later time delays when the NC lattice temperature reaches some critical value. For a better understanding of this process the measurement of Au$_{923}$ on Si-N has been repeated for different fluences yielding similar time-evolutions (**fig. 5.c**). To estimate the maximum increase of width, we have averaged the experimental values from 14 to 30 ps (blue datapoints). The standard deviation at the same time interval have been used to estimate the corresponding error-bars. The results of this procedure are plotted as a function of the estimated, incident fluence, in **figure 5.d**. The fitted function (red solid line) has the form $a \cdot F^n$, where F is the fluence and n=2.4.

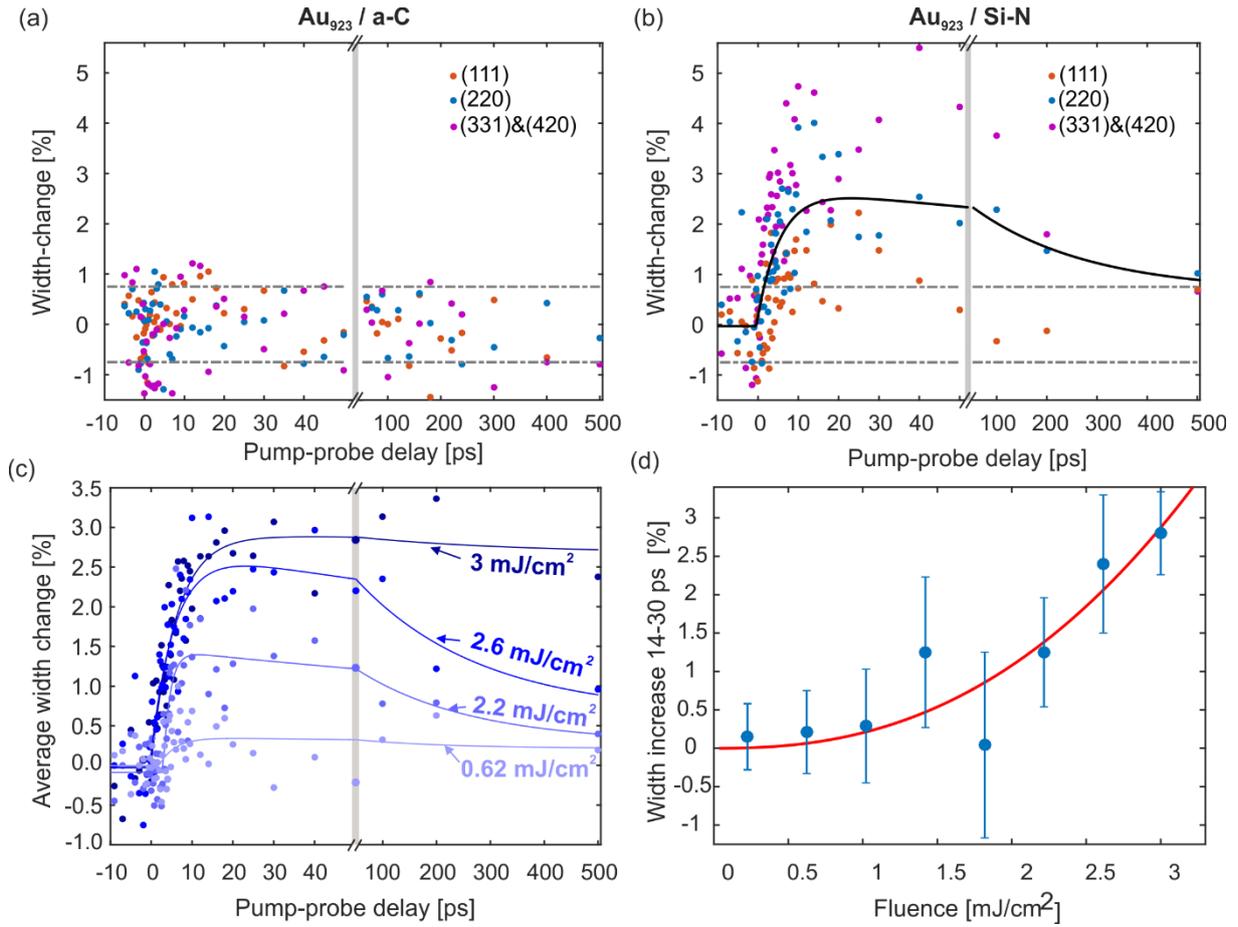

**Figure 5: Ultrafast evolution and fluence dependence of diffraction peak broadening.** (a) Time-dependent relative diffraction peak width (full width at half maximum, FWHM) for three $Au_{923}$ peaks on a-C. (b) Same for $Au_{923}$ on Si-N. The black solid line is a bi-exponential fit of the average width increase. (c) Average peak broadening for $Au_{923}$ on Si-N (blue data points) and bi-exponential fittings (blue solid lines) for different fluences. (d) Average peak broadening as a function of the incident fluence for $Au_{923}$ on Si-N (blue points) and error-bars estimated from the standard deviation for time delays between 14 and 30 ps.

As illustrated in **fig. 3.c**, an increased width of diffraction peaks is an indicator of a deterioration of crystallinity, for instance, through the formation of a non-crystalline outer shell or through the development of inhomogeneous strain. The underlying process must by non-thermal and driven by electrons, this is because: (i) the peak broadening is fast and starts rising immediately after photo-excitation with a time-constant similar with that of heat flow to the Au lattice, (ii) the energy content of the NCs in all measurements is below the reported threshold for initiation of surface pre-melting effects at this size-regime[14] (T~800 K) and (iii) the effect takes place only in the presence of hot electrons (**fig. 5.b**) and remains unseen for slow, vibrational coupling (**fig. 5.a**). Finally, the non-linear dependence on the incident fluence (**fig. 5.d**) suggests that the participating atoms have to overcome an energy barrier.

**Origin of hot-electron-driven peak broadening:** There are two phenomena that possibly account for the observed peak broadening: (1) activation of atomic diffusion, leading to a liquid-like layer on the surface of NCs and a reduction of the size of the crystalline core, (2) preferential coupling of electrons to collective vibrations (breathing and toroidal modes) that cause incoherent fluctuations of the NC size and shape. Breathing and toroidal modes can cause time-evolving variations on the atomic displacements across the nanostructure volume resulting in inhomogeneous strain.[63] Coherent oscillations might not produce a coherent signal due to the NCs size dispersion[64] but they can give rise to an averaged peak broadening. At this size regime, the period of collective vibrations is in the order of 1 ps or lower.[8] Due to the small size of the NCs, the laser excitation will results in a homogeneous hot electron distribution within tens of femtoseconds. This is fundamentally different compared to bulk-like samples or large nanoparticles. We therefore exclude inhomogeneous electron excitation within the NCs as cause of inhomogeneous strain[65] or structural changes induced by the hot electron blast force.[28]

We performed a size-strain analysis by comparing the broadening dynamics of two diffraction peaks,[66] see supplementary information. The analysis of the photo-induced changes of the peak widths suggests that approximately 90% of the effect is caused by a reduction of the diameter of the crystallite, *i.e.*, by approximately 1.7%. This change in size corresponds to the removal of (50±20) atoms from the crystalline core, corresponding to ~14% of the surface atoms (for $Au_{923}$, the surface layer contains 362 atoms, *i.e.*, the difference with the previous magic number, $Au_{561}$). The measurement with the highest fluence in **figure 5.c**, shows peak broadening for pump-probe delays up to 0.5 ns. The slow recovery of peak width is suggestive of slow recrystallization of a liquid-like surface layer, while collective vibrations are expected to be damped on shorter time scales due to the small size of NCs and interactions with the substrate.[8] While inhomogeneous strain might facilitate surface disordering as discussed below, there is additional indications why strain is not the main cause for the peak broadening itself. First, the timescale of coherent acoustic vibrations (~1 ps)[8] is distinctively faster than the observed peak broadening. Second, strain induced by coherent oscillations is expected to have a linear dependence on the vibrational amplitude and subsequently a sublinear dependence on the energy input. Thus, contributions from strain cannot account for the strongly non-linear increase of broadening with the excitation fluence in **figure 5**. Instead, the relevant lifetime, close to that of electron-phonon coupling, and the electron temperature dependence (Sup. Info. section 8) suggest that the effect is triggered by hot electrons.

Our observations and considerations suggest that the dynamic diffraction peak broadening contains contribution from surface disordering of the Au NCs which only occurs in the presence of hot electrons under pronounced electron-lattice non-equilibrium. This conclusion agrees with previous experiments. Taylor *et al.*[67] have shown that laser-irradiated Au nanorods transform into nanospheres below the melting point through diffusion of surface atoms. Clark *et al.*[25] have directly visualized the formation of such a liquid, surface layer on a laser-excited Au nanoparticle using coherent diffractive X-ray imaging. In this experiment, surface disordering occurred at a lower temperature than the melting point of bulk Au and recrystallization was slower than 500 ps.

The activation of surface diffusion of Au NCs in thermal equilibrium has been studied by *in situ* TEM measurements.[68] The range of values for the activation energy reported in that work, incorporated into the transition state theory of Eyring,[69] cannot explain the observed dynamics and the dependence of the peak broadening on the substrate material.

Our experimental findings have some similarities with the ultrafast desorption or diffusion of adsorbates on photo-excited metallic surfaces. Previous experimental[70-72] and theoretical work[73,74] has shown that hot electrons can enhance the diffusion of adsorbed molecular species on metallic surfaces due to effective excitation of adsorbate vibrations through electronic friction.[75,76] So far, self-diffusion of metal adatoms has not been taken into account in this context because the effects of electronic friction are considered less important as the atomic mass increases. However, it is known that the potential energy surface of Au changes at high electronic temperatures.[77,78] The question if hot electrons have an effect also on the surface atoms of the metal itself is largely unexplored, partially because ultrafast spectroscopic techniques do not provide direct information on the structural properties and suffer from low contrast of signatures from surface atoms compared to the bulk. Our approach of studying the structural dynamics of nanostructures with large surface-to-volume ratio with FED, however, provides sufficient sensitivity to the surface crystallinity. To explain the observed disordering of metal atoms by electronic excitations, we consider the following scenario: NCs exhibit a high density of surface states near the Fermi level, for instance Shockley surface states or adsorbate-induced states. On Au nano-facets, surface states have been shown to be spatially inhomogeneous through scanning tunneling spectroscopy.[79] Laser excitation leads to a (de-)population of electrons (below) above the Fermi level. We expect that this population redistribution results in fluctuating microscopic electric fields, both in and out of the surface plane. This process will transfer energy from the electrons to vibrations of the surface atoms, in addition to conventional electron-phonon coupling. In addition, the fluctuation in surface state occupation will result in fluctuating potential energy of the surface atoms. Through these mechanisms, surface atoms accumulate vibrational excitation and eventually get removed from their equilibrium positions. Since the proposed mechanism involves electronic surface states, the exact microscopic mechanism might depend on the facet and adsorbate coverage.

In support of this hypothesis, we note that: (i) metal surfaces exhibit enhanced atomic diffusion and reconstruction when surface-atoms get polarized by a static electric field,[80,81] (ii) spatially confined surface states are affecting the diffusion of metal adatoms on metal surfaces,[82,83] (iii) scattering by surface imperfections is a major decay-channel for surface states,[84,85] and (iv) strain of metal surfaces modifies the diffusion barrier.[86] The proposed scenario has analogies to the DIMET mechanism (desorption induced by multiple electronic transitions),[87] with electronic transitions occurring between bulk and surface states, surface atoms playing the role of adsorbates, and diffusion being the resulting process instead of desorption. A final remark is that strong electric fields on the NC surface can also be created by image-potential states or thermionic emission. However, these transitions have much lower probability due to their high energy.[88]

**Hot electrons in surface chemical reactions:** We expect our findings to be of relevance for a range of structural dynamics in nanostructures. For example, laser-excited Pd nanostructures,[89] Au/VO$_2$ interfaces[90] and Au/MoS$_2$ heterostructures[91] have been shown to exhibit structural transformations that in many cases have been attributed to hot electrons. Hot electrons, however, do not only occur after optical excitation with femtosecond laser pulses but also during surface chemical reactions.[92] Past studies have shown that the catalytic activity of Au NCs is accompanied by changes of their morphology,[93] more precisely sintering of the NCs, which degrades the catalytic activity. Yang *et al.*[94] have shown that during catalytically-activated CO oxidation, the activation energy for sintering of Au NCs shows a 28-fold decrease. It was suggested that the detachment of surface atoms takes place due to interactions with excited electrons generated by CO oxidation. These highly excited electrons reside on surface states of the NC, which, as discussed before, can lead to surface atom diffusion.

**SUMMARY:**

The ability to prepare 2D distributions of NCs with narrow distribution of sizes and structural allotropes, supported on various thin-films, minimizes the uncertainty of structural characterization and allows FED to track the ultrafast atomic motions after photo-excitation.[22,95] We employed light-absorbing substrates (a-C) as well as transparent substrates (Si-N) to modify the pathway of ultrafast heat flow and for comparing the structural changes driven by hot electrons and hot phonons. Measurements on both substrates yielded a reduced thermal expansion coefficient compared to bulk Au. Static measurements of $Au_{923}$ on graphene have revealed that the lattice is compressed, even at equilibrium, due to surface tension. The next observable that has been compared between the two substrates was the peak-broadening after excitation, which reflects atomic motions that reduce long-range order. This effect takes place only in the presence of hot electrons and arises from atomic disordering of the NC surfaces. We propose a mechanism leading to an enhanced excitation of surface vibrations and fluctuations of the surface potential. Our findings strongly support the general notion that hot electrons trigger surface atom diffusion in Au NCs.

The proposed mechanism of peak-broadening due to surface-diffusion requires that the potential energy surface is modified at high electronic temperatures. In this scenario, optical excitation with the laser pulse leads to multiple electronic transitions that involve surface states, each of them characterized by an inhomogeneous, polarizing, electric field. As a result the potential energy surface is fluctuating. Through this mechanism, surface atoms accumulate vibrational excitation and eventually get removed from their equilibrium positions. Excitation-induced strain may contribute by modulation of the surface atom diffusion barrier. Although the exact mechanism is still to be verified, the possibility that electronic excitations facilitate surface diffusion, can be an essential part of the morphological changes of NCs induced by surface chemical reactions or laser irradiation.

**METHODS:**

**Size-selected synthesis:** Size-selected Au nanoclusters were produced using a magnetron sputtering gas aggregation cluster source[4] with a lateral time-of-flight mass filter.[96] A mass resolution of M/ΔM= 20 was applied to give clusters with 923±23 atoms. The clusters were deposited on either copper TEM-grids coated with a thin film of a-C (20 nm thick), silicon TEM-grids with an array of Si-N windows (100 μm wide and 10 nm thick) and few-layer graphene (used as a reference-material to investigate static lattice-compression due to surface-tension). The $Au_{923}$ NCs were deposited in the 'soft-landing' regime (<2 eV/atom) to prevent fragmentation and maintain the cluster structure.[5] The deposition density was 8 clusters per 100 $nm^2$ and all formation conditions were kept the same so as not to change the relative proportions of structural isomers.[3] To immobilize the clusters, defects were created prior to cluster deposition, by exposure of the substrate to a high energy Ar+ beam (1500 V).

After deposition, the samples on conducting substrates were examined using a 200 keV JEOL JEM 2100F STEM (scanning transmission electron microscope) with a spherical aberration corrector (CEOS). A HAADF detector with inner collection angle of 62 mrad was employed for z-contrast imaging of the samples; in this mode, the intensity is linearly proportional to the number of atoms.[97] This enabled the 3D atomic structure of the clusters to be determined[97] and for the size distribution to be calculated using the size-selected $Au_{923}$ clusters as a mass-balances.[98]

**FED experiments:** Time-resolved measurements have been carried out with FED operated with a repetition rate of 1 kHz. Individual electron bunches contained a few thousand electrons. The diffractometer is characterized by a highly compact design and the transverse coherence of the electron beam is larger than the diameter of the Au NCs under investigation.[33] For each time-delay, 40,000 diffraction patterns have been integrated for $Au_{923}$ on a-C and 20,000 to 40,000 for the fluence dependence measurements on Si-N. The acceleration voltage of the electrons was kept at 93 keV and the estimated time-resolution was in the order of ~300 fs. The substrate's contribution to the diffraction patterns has been taken into account by recording the diffraction pattern of a bare substrate under the same experimental conditions.

**ACKNOWLEDGMENT**

We thank Alexander Paarmann, Alexey Melnikov, Martin Wolf and Peter Saalfrank for useful comments and discussions. This project has received funding from the Max Planck Society and from the European Research Council (ERC) under the European Union's Horizon 2020 research and innovation program (grant agreement number ERC-2015-CoG-682843). R.B. acknowledges funding from the Alexander von Humboldt Foundation and A.S. from the Brazilian National Council of Technological and Scientific Development (CNPq).

**SUPPORTING INFORMATION**

Supplemental material (PDF) includes: (i) illustration of the background subtraction before fitting the time-resolved diffraction patterns, (ii) static measurements of Au NCs on graphene to extract lattice compression at equilibrium, (iii) additional measurements of the electron-phonon coupling constant for $Au_{923}$ on Si-N, (iv) measurement of the electron-phonon coupling constant for bulk-like, free-standing, Au films, (v) peak broadening as a function of the maximum electronic temperature. (vi) a discussion on the effect of photo-induced charging at higher laser fluences and its effect on the FED measurements, (vii) a test measurement at different sample and experimental conditions, (viii) additional analysis of the anharmonicity of $Au_{923}$ NCs and (ix) fluence dependence measurements of peak broadening for $Au_{923}$ on a-C.

Gold Nanoparticles Studied by Ultrafast Electron Nanocrystallography. *Nano Lett.* **2007**, *7*, 1290–1296.

(23) Plech, A.; Kotaidis, V.; Grésillon, S.; Dahmen, C.; Von Plessen, G. Laser-Induced Heating and Melting of Gold Nanoparticles Studied by Time-Resolved X-Ray Scattering. *Phys. Rev. B* **2004**, *70*, 195423.

(24) Mancini, G. F.; Latychevskaia, T.; Pennacchio, F.; Reguera, J.; Stellacci, F.; Carbone, F. Order/Disorder Dynamics in a Dodecanethiol-Capped Gold Nanoparticles Supracrystal by Small-Angle Ultrafast Electron Diffraction. *Nano Lett.* **2016**, *16*, 2705–2713.

(25) Clark , J.N; Beitra, L.; Xiong, G.; Fritz, D.M., Lemke; H.T.; Zhu, D.; Chollet, M.; Williams, G.J; Messerschmidt, M.M.; Abbey, B.; Harder, R.J.; Korsunsky, A.M.; Wark, J.S.; Reis, D.A.; Robinson, I.K.  Imaging Transient Melting of a Nanocrystal Using an X-ray Laser. *PNAS* **2015**, *112*, 7444–7448.

(26) Liang, W.; Schäfer, S.; Zewail, A. H. Ultrafast Electron Crystallography of Heterogeneous Structures: Gold-Graphene Bilayer and Ligand-Encapsulated Nanogold on Graphene. *Chem. Phys. Lett.* **2012**, *542*, 8–12.

(27) Sokolowski-Tinten, K.; Shen, X.; Zheng, Q.; Chase, T.; Coffee, R.; Jerman, M.; Li, R. K.; Ligges, M.; Makasyuk, I.; Mo, M.;Reid, A.H.; Rethfeld, B.; Vecchione, T.; Weathersby, S.P.; Dürr, H.A.; Wang, X.J. Electron-Lattice Energy Relaxation in Laser-Excited Thin-Film Au-Insulator Heterostructures Studied by Ultrafast MeV Electron Diffraction. *Struct. Dyn.* **2017**, *4*, 054501.

(28) Esmail, A. R.; Bugayev, A.; Elsayed-Ali, H. E. Electron Diffraction Studies of Structural Dynamics of Bismuth Nanoparticles. *J. Phys. Chem. C* **2013**, *117*, 9035–9041.

(29) Vanacore, G. M.; Hu, J.; Liang, W.; Bietti, S.; Sanguinetti, S.; Zewail, A. H. Diffraction
27

# Supplementary Information:

# Ultrafast Heat Flow in Heterostructures of Au Nanoclusters on Thin-Films: Atomic-Disorder Induced by Hot Electrons

*Thomas Vasileiadis†, Lutz Waldecker†, Dawn Foster‡, Alessandra Da Silva‡, Daniela Zahn†, Roman Bertoni†, Richard E. Palmer§, and Ralph Ernstorfer†*

† Fritz-Haber-Institut, Faradayweg 4-6, 14195 Berlin, Germany

‡Nanoscale Physics Research Laboratory, School of Physics and Astronomy, University of Birmingham, Edgbaston, Birmingham B15 2TT, United Kingdom

§College of Engineering, Swansea University, Bay Campus, Fabian Way, Swansea SA1 8EN, United Kingdom

1) **Background subtraction**

**Figure S1** shows the fitting procedure of $Au_{923}$ NCs on Si-N. Fitting the raw data (blue, solid line) starts with the subtraction of a background (red, solid line). This background contains the diffraction pattern of a bare substrate recorded at the same conditions (brown, dash-dot) and a lorentzian function located at zero angle for the extended wing of the zero-order peak (green, dash-dot). The final curve (black) is then fitted with a smooth background (third-order polynomial) in the flat regions between the peaks that corresponds to inelastic scattering of electrons from phonons. The signal in these flat areas is used to record the increase of the inelastic scattering background (black curve, **figure 2.a** and **2.b** in the main article) The remaining peaks are fitted with pseudo-Voigt peak-profiles. After this data-treatment the distance from the zero order (in pixels) is transformed into scattering vector using literature values for the crystal structure of Au.

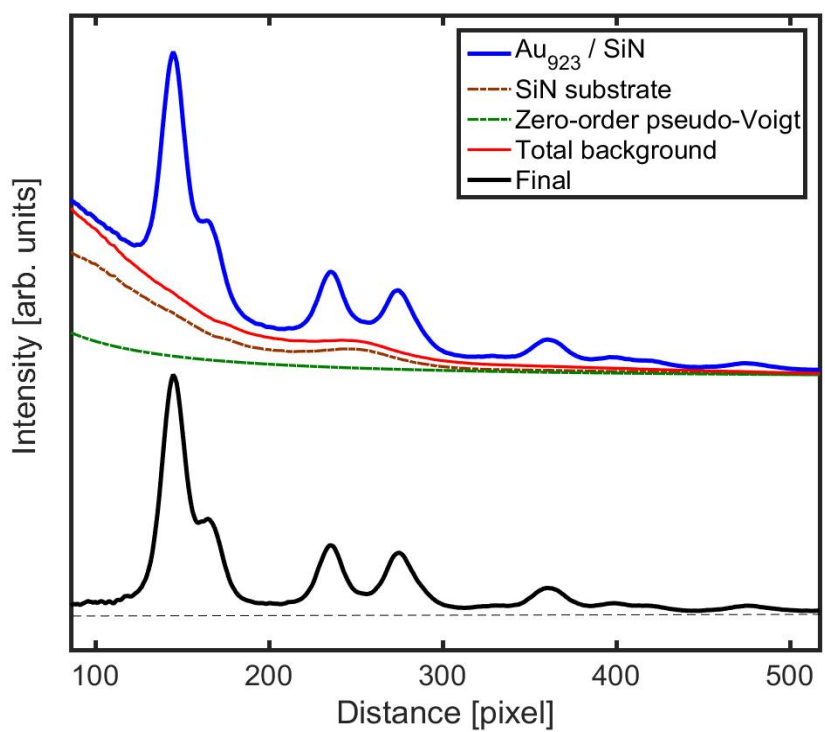

**Figure S1: Demonstration of background subtraction procedure for $Au_{923}$ on Si-N.**

2) **Extraction of time-constants**

The characteristic time-scale of a photo-induced change can be quantified, approximately, through the relationship:

$$F(t) = c + \Theta(t - t_o) \cdot \sum_{i=1}^{n} A_i \left(1 - \exp\left[-\frac{t - t_o}{\tau_i}\right]\right)$$

where $F(t)$ is the time-evolving observable, $c$ is its value before photo-excitation, and the second term is an approximation of its change after photo-excitation. The number of photo-induced processes is given by $n$ (for bi-exponential $n = 2$). Each process is changing the observable by $A_i$ with a time-constant $\tau_i$. The step function $\Theta(t)$ expresses that photo-excitation happens at $t = t_o$. The time-resolution of the setup is taken into account through a convolution of the above function with a Gaussian.

3) **Lattice compression of $Au_{923}$ on few-layer-graphene**

**Figure S2** depicts a specific region of the radial-averages of $Au_{923}$ NCs / few-layer-graphene heterostructures (data-points), that has been used to estimate volume-changes (compared to bulk Au) of $Au_{923}$ at equilibrium. The recorded diffraction pattern of bare graphene substrates is also shown for comparison (black line). Having graphene, at the exact same area with NCs, eliminates the uncertainties introduced by the electron-optics. The two peaks, marked by arrows, are the (220) peak of Au (with an inter-plane spacing of $D_{220}$ = 1.4391 Å for bulk Au) and the (110) peak of graphene with inter-plane spacing of $D_{110}$ = 1.2280 Å. In these experiment the (220) peak appeared at a larger scattering vector compared to bulk Au, meaning that the $Au_{923}$ nano-lattices are compressed. This compression can arise by enhanced surface-tension due to high surface-to-volume ratio. More details can be found in the main article. We have recorded and fitted separately 11 diffraction patterns with 5 s exposure time each (meaning that 55,000 electron pulses got diffracted by the sample in total). With this procedure we find that the nano-lattice of Au$_{923}$ NCs is compressed at equilibrium by (0.63±0.05)% compared to bulk Au.

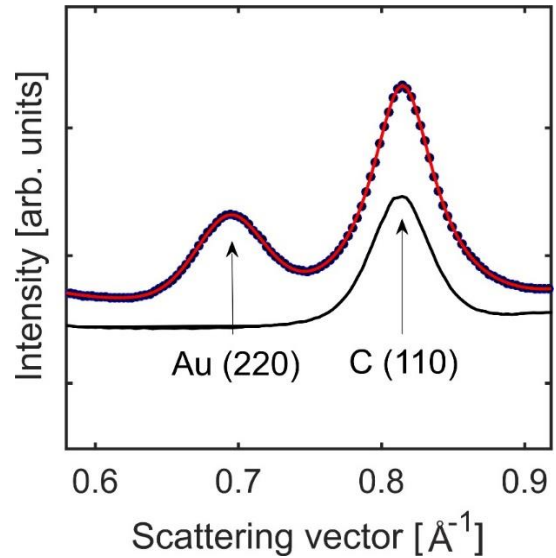

**Figure S2: Lattice-compression of Au$_{923}$ NCs, at room temperature, measured with few-layer graphene substrate as a reference.** Static radial averages of Au$_{923}$/graphene heterostructures (data-points), fitting using pseudo-Voigt peak-profiles (red line) and radial average of the bare graphene-substrates (black line).

4) **Heat capacities for the model of energy flow**

For the electronic heat capacity of Au$_{923}$ we have assumed that it is similar with bulk Au. There are several reasons to expect that at this size-regime (D~3 nm) electrons will not be significantly affected from spatial confinement. A first aspect, is that Coulomb screening will reduce surface effects. For Au the Thomas-Fermi screening length ($r_{TF} \sim 1.6$ Å) is much smaller than the dimensions of Au$_{923}$, and hence the fraction of electrons that is affected by the surface, will be limited. A second aspect, is that any discretization of the electronic density of states is smoothed out because of thermal broadening. If the electronic density of states close to the Fermi energy acquires gaps (due to the finite amount of atoms) of equal spacing then the critical size for major confinement effects at room temperature is $E_F/k_BT \sim 200$ atoms.

Regarding the lattice, we extrapolate the lattice heat capacity of Au NCs at low temperatures (Sauceda *et al.*, ref. 53 in main article), to the desired range using the vibrational density of states $F(\omega)$ (Sauceda *et al.*, ref. 54 in main article) according to:

$$C_l = \int \frac{\partial n_B(\hbar\omega, T_l)}{\partial T_l} F(\omega)\hbar\omega \, d\omega$$

where $n_B(\hbar\omega, T_l)$ is the Bose-Einstein distribution function. The heat capacity at low temperatures and the quantity $F(\omega)$ have been calculated by averaging the vibrational density of states of the various structural allotropes of similar size clusters, weighted by their relative abundances obtained from HR-TEM measurements. The results are shown in **figure S3**.

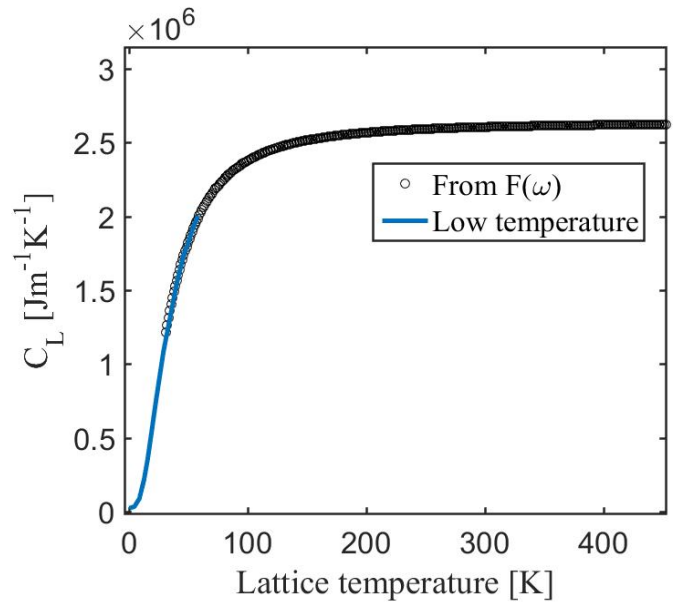

**Figure S3: The heat capacity of Au NCs used for fitting the lattice temperature evolution.**

5) **Electron-phonon coupling measurements at different fluences**

We have measured the lattice temperature evolution from four different measurements at fluences: 3, 2.6, 2.2 and 1.8 mJ/cm$^2$ and the fittings together with the experimental data are shown in **figures S4 (a) to (d)** respectively. The fitted values for $G_{e-ph}$ are: (a) $2.079 \cdot 10^{16}$ W/m$^3$K, (b) $2.31 \cdot 10^{16}$ W/m$^3$K, (c) $1.79 \cdot 10^{16}$ W/m$^3$K and (d) $1.86 \cdot 10^{16}$ W/m$^3$K.

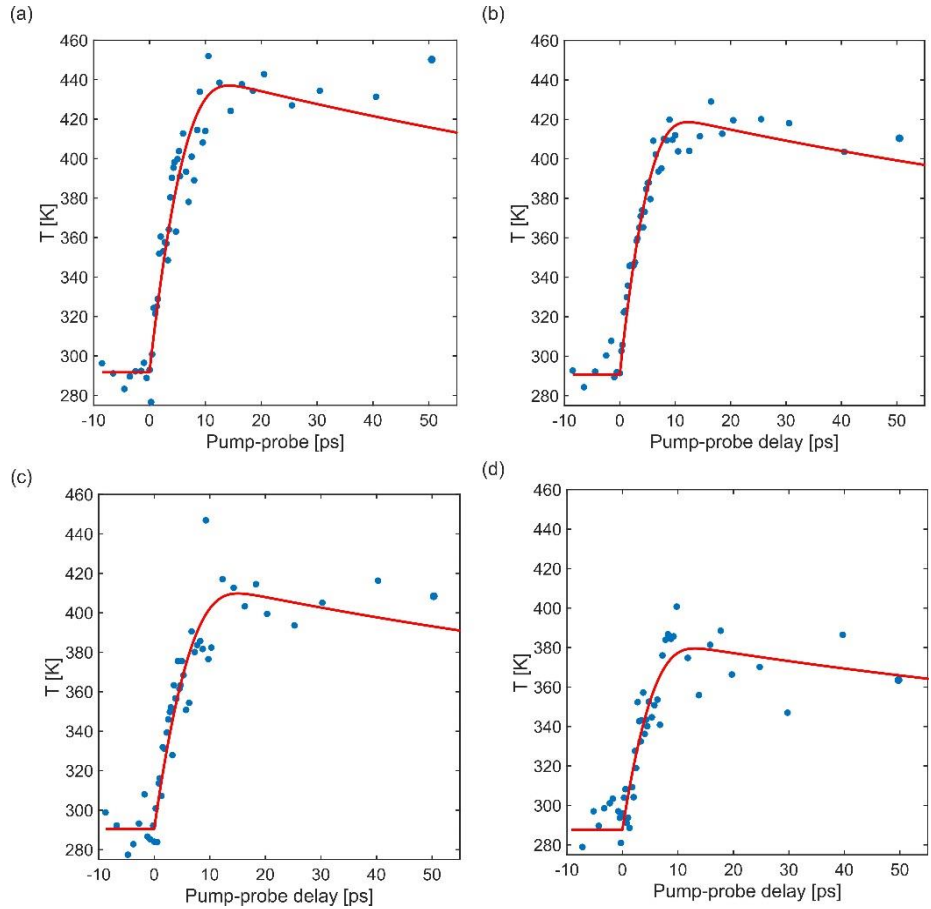

**Figure S4: Experimental lattice temperature evolutions shown at short delays for four different fluences and the corresponding fittings using the model of energy flow in heterostructures.**

The lattice temperature in **figure S4** is calculated with the Debye-Waller effect observed for the peaks (111), (220) and the average of (331) and (420). As an additional check we have used the lattice heat capacity and Debye-Waller factor of bulk Au to fit the data of the Au$_{923}$ NCs. With this simplification the extracted electron-phonon coupling constant is again lower than bulk Au and equal with $(2.1\pm0.3) \cdot 10^{16}$ W/m$^3$K.

## 6) Lattice-dynamics of bulk-like, free-standing, thin-films of Au

To ensure the capabilities of the used apparatus and the analysis we have conducted time-resolved experiments on bulk-like, free-standing, thin-films of Au with 8 nm thickness. The fitted, time-evolving quantity is the lattice-temperature estimated from the Debye-Waller effect. The electron-phonon coupling constant, based on the two-temperature-model (TTM) was found equal with $G = 0.27 \cdot 10^{17} \, W/m^3 K$. It should be noted that at this excitation level (the maximum electronic temperature is 2950 K), G is nearly constant.

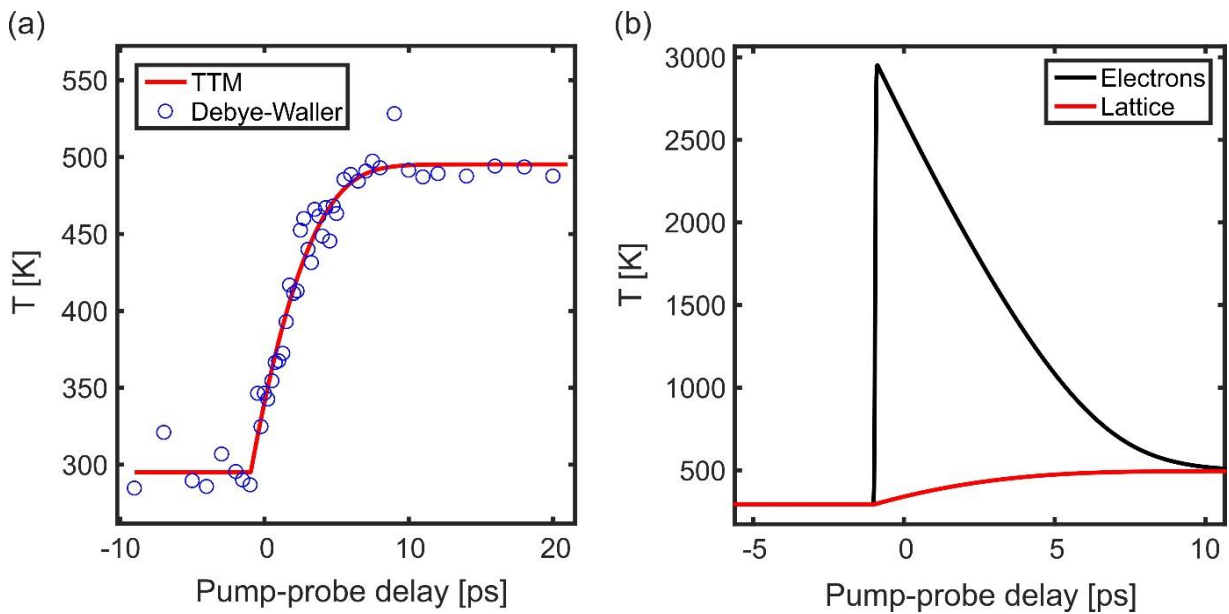

**Figure S5: Evolution of lattice temperature for bulk-like 8 nm thin-films of Au and modelling using the TTM.** (a) Experimental evolution of lattice-temperature based on the assumptions of the Debye-Waller effect (blue data-points) and fitting using the TTM (red, solid line). (b) Electronic (black, solid line) and lattice temperature (red, solid line) based on the TTM.

## 7) Size-strain analysis

The Williamson-Hall method (reference 66 in the main article) have been applied in the measurement with the highest fluence and peak broadening to estimate the contributions of crystallinity reduction and inhomogeneous strain increase. The width of diffraction peaks β is given by:

$$\beta = \frac{K\lambda}{D \cos \Theta} + 4\varepsilon \tan \Theta$$

where Θ is the angle of each diffraction peak (time-evolving due to lattice expansion), $D$ is the crystalline size of the NC, $\varepsilon$ is the inhomogeneous strain and $K$ is a constant close to unity. The first term is broadening related with the limited size of the NCs and the second term, broadening due to inhomogeneous strain. The quantity $\beta \cos \Theta$ is plotted versus $\sin \Theta$ for every peak and for every time delay and then the slope and the y-intercept are extracted. The slope is $A = 4\varepsilon$ and thus it is proportional to inhomogeneous strain and the y-intercept is $B = K\lambda/D$ and thus it is inversely proportional to the NCs size. The diffraction peak (331)&(420) is not appropriate for the above analysis because it is double and its width is affected by expansion. The plot is shown in **figure S6** for the (111) and (220) diffraction peaks. For time delays between 12 and 30 ps, the slope is slightly increasing compared to negative time delays (**fig. S6**) and this indicates a slight increase of inhomogeneous strain. The y-intercept is also increasing from $(8.7\pm0.1)\cdot10^{-4}$ to $(8.8\pm0.1)\cdot10^{-4}$. This corresponds to a decreased radius of the NCs by 1.7%. Changes of the size account for approximately 90% of the overall effect. Starting from a total of 923 crystalline atoms on a sphere the decrease in the radius corresponds to (50±20) surface atoms leaving the crystalline core and becoming mobile.

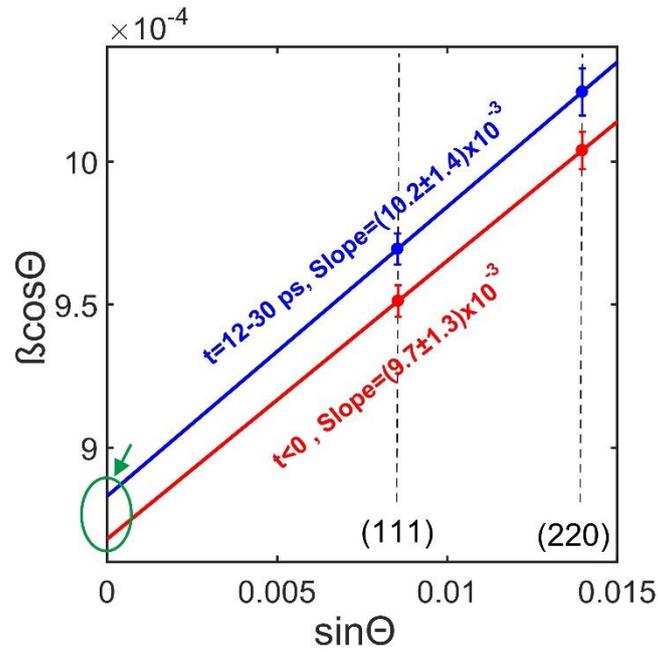

**Figure S6: Williamson-Hall plot for the (111) and (220) diffraction peaks at negative and positive time delays.** The data-points are averaged for negative time delay (red) and for 12-30 ps (blue). The error-bars are estimated from the standard deviation of the data at the same intervals. The error-bars on the x-axis are smaller than the data-points. The increase of the y-intercept is indicated with a green arrow.

8) **Maximum electronic temperature and peak broadening**

For all measurements the model of heat flow in heterostructures has been to estimate the time-evolving heat content of the electronic subsystem. **Figure S7** shows the maximum peak-broadening (average peak broadening at 14-30 ps) versus the maximum electronic temperature estimated from the model. Peak broadening becomes significant when the maximum electronic temperature exceeds ~3000 K.

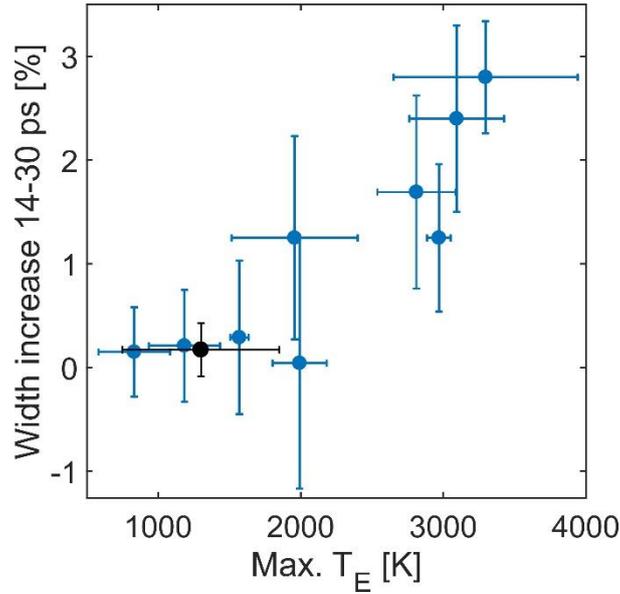

**Figure S7: Maximum peak broadening versus maximum electronic temperature.** The blue data-points are a fluence dependence measurement of size-selected $Au_{923}$ NCs on Si-N. The black data-point is the measurement on a-C presented in the main article. The error-bars for the width increase are the standard deviation of the data from 14-30 ps. The error-bars for the maximum electronic temperature are estimated using various values for the temperature dependence of the Debye-Waller factor and the heat capacity.

9) **Effect of photo-ejected electrons**

Besides real space atomic motions, the diffracted electron bunches can also be affected by electrostatic deflections due to photo-ejected electrons (although their effect is expected to be small for the incident laser fluences and the diffracted electron energies used here). Even if such photo-electrons exist, they cannot explain the observed increase of the width of diffraction peaks. The reason is that the diffracted electrons have all passed through the interior of one NC and thus they should mostly be affected by the positive charge that remains after photo-ejection of electrons. Hence, the deflection should be maximum after the laser pulse gets absorbed and not several picoseconds after.

## 10) Ultrafast shrinking of diffraction ring radius at high laser fluence

At high laser fluence the shrinking of diffraction peaks (expansion) shows an ultrafast, time-resolution limited component (**fig. S8**). This signal can originate on an ultrafast expansion of Au NCs but it can also contain contributions from deflections of the electron beam due to photo-charges, as discussed above.

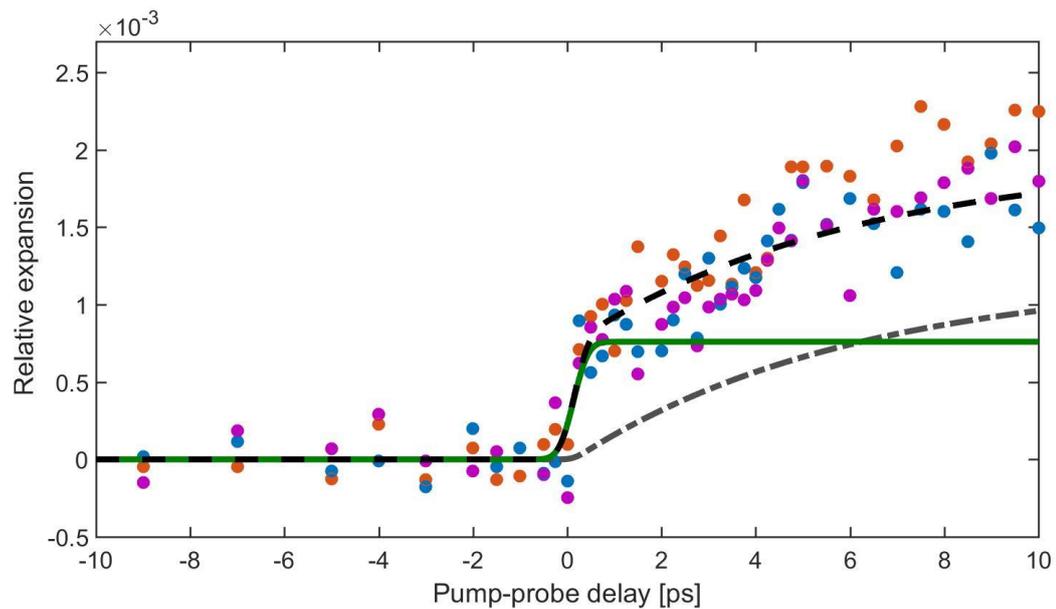

**Figure S8: Ultrafast change of diffraction ring radius at short time delays.** Relative expansion for 3 mJ/cm$^2$ and for three different peaks: (111), (220) and the average of (331) and (420) (orange, blue and magenta data-points respectively). The dashed black line is a fitting with an exponential growth (dashed-dot grey line) plus a process faster than the time-resolution of the instrument (green solid line).

## 11) Additional FED measurement on Au$_{923}$ / Si-N heterostructures

We have performed one additional measurement on a different sample of Au$_{923}$ / Si-N heterostructures. Although the statistics of this measurement were lower, it is in general agreement with our results. The lattice temperature evolution has the familiar dynamics, a fast increase followed by a slow decay, and the average width is increasing after excitation (**fig. S9**). The average width broadening for (111), (220), (331) and (420) and for pump-probe delays between 14 and 30 picoseconds is (1.7±0.9)% (maximum lattice temperature is ~390 K). This measurement was carried out at a lower accelerating voltage of 62 kV instead of 93 kV (main manuscript). Considering the degree of excitation (**fig. S9.a**) the width increase is in the expected range.

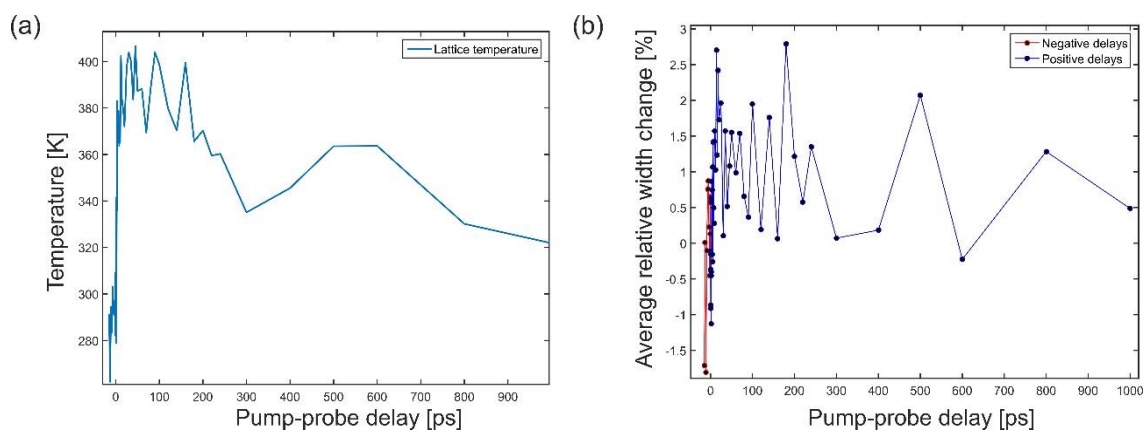

**Figure S9: Measurement on a different sample of Au$_{923}$ on Si-N with a lower gun voltage (63 kV).** (a) Lattice temperature from the Debye-Waller effect and (b) relative width increase averaged for the (111), (220) and (331) and (420) peaks of Au$_{923}$ NCs.

**12) Anharmonicity of Au$_{923}$ NCs**

In the main article we have presented measurements of the thermal expansion coefficient for Au$_{923}$ NCs. An alternative way to quantify the anharmonic properties of NCs and to compare them with bulk Au is to measure the proportionality factor (A) between the atomic mean square displacement (MSD) and the observed expansion: $\delta L/L = A \cdot [<u^2>_{(t)} - <u^2>_{(t<0)}]$. The advantage is that both quantities are directly measured by FED. **Figure S10** shows the mean displacement plotted versus mean-suare-displacement from different measurements and fittings. In the case of bulk Au, calculations give A= 0.175 Å$^{-2}$. For Au$_{923}$ on Si-N this value is reduced by (8.7±1.6)% while for Au$_{923}$ on a-C the reduction is (21.6±1.9)% compared to bulk. These values are in agreement with the reduction of anharmonicity stated in the main manuscript. The difference between the two substrates is expected. On the one hand, lattice-expansion on Si-N coexists with non-thermal phonon distributions and atomic disordering. On the other hand, lattice-heating on a-C is slower and resembles measurements close to thermodynamic equilibrium.

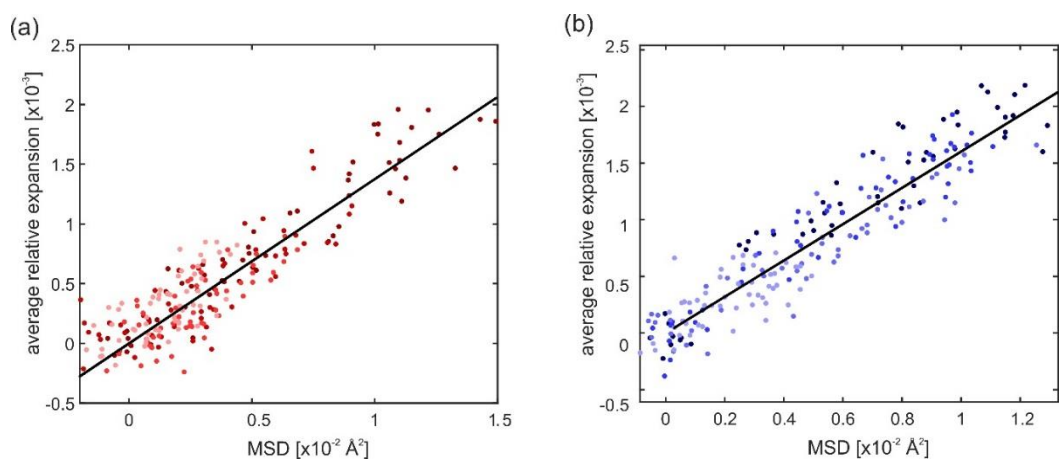

**Figure S10: Anharmonicity of Au$_{923}$ NCs on (a) a-C and on (b) Si-N.** The average expansion plotted as a function of the increase of the atomic mean-square-displacement (MSD) for multiple measurements. Only the positive time-delays have been fitted.

### 13) Peak broadening for A$_{923}$ / a-C

Fluence dependence measurements have been carried out also for Au$_{923}$ on a-C. The measurement with the highest fluence is shown in the main manuscript. **Figure S11** shows the average change of width for the (111), (220) and (331) and (420). The width increase, averaged over 14-30 ps, was always zero.

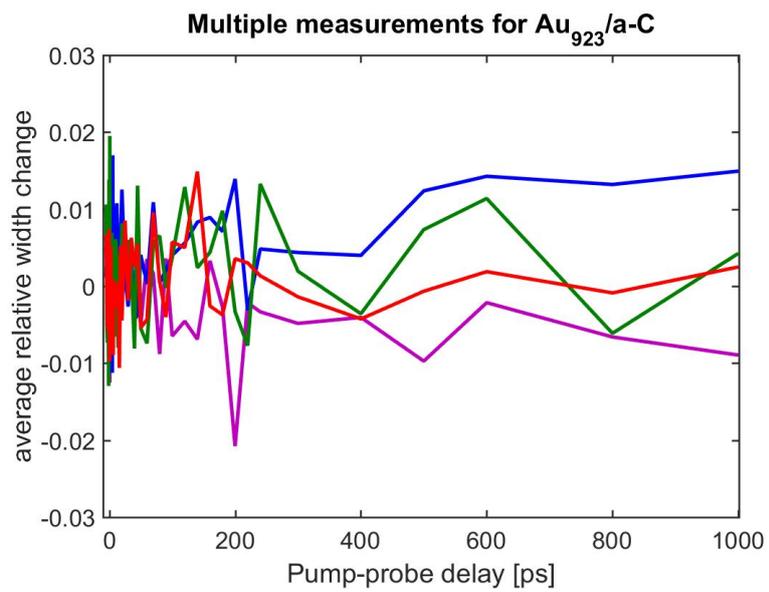

**Figure S11: Multiple time-resolved measurements of peak width for Au$_{923}$ on a-C.** The relative width change averaged over the (111), (220) and (331) and (420) for different fluences (different colors).